\documentclass[aps,prd,preprint,a4paper]{revtex4}
\usepackage{epsfig}
\begin{document}
\title{On Dyson-Schwinger equations and the number of fermion families}
\author{Davor Palle}
\affiliation{
Zavod za teorijsku fiziku, Institut Rugjer Bo\v skovi\' c \\
Bijeni\v cka cesta 54, 10001 Zagreb, Croatia}

\date{November 25, 2009}

\begin{abstract}
We study Dyson-Schwinger equations for propagators of Dirac fermions
interacting with a massive gauge boson in the ladder approximation.
The equations have the form of the coupled nonlinear integral Fredholm
equations of the second kind in the spacelike domain.
The solutions in the timelike domain are completely defined by 
evaluations of integrals of the spacelike domain solutions.
We solve the equations and analyze the behavior of solutions on 
the mass of the gauge boson, the coupling constant, and the ultraviolet cutoff.
We find that there are at least two solutions for the fixed gauge
boson mass, coupling, and the ultraviolet cutoff, thus there are
at least two fermion families.
The zero-node solution represents the heaviest Dirac fermion state,
while the one-node solution is the lighter one. The mass gap between
the two families is of the order of magnitude observed in nature.
\end{abstract}

\maketitle

\vspace{2ex}

\section{Introduction and motivation}

There is a common belief in particle physics that the Higgs mechanism
resolves the problem of generating the masses of particles, although
the masses of gauge bosons and fermions are fixed by couplings
to Higgs scalars that are completely free parameters.
The observed pattern of fermion and gauge boson masses
should stimulate a search for a better symmetry breaking mechanism.
The example of such a symmetry breaking mechanism is the proposed
principle of noncontractible space \cite{Palle1}.

It is assumed that the origin of the broken conformal, discrete, and
gauge symmetries in particle physics is hidden in the character
of the physical space, namely its noncontractibility.
The appearance of the chirally asymmetric coupling of SU(2) gauge
bosons to leptons and quarks, the appearance of the massive SU(2)
gauge bosons and very heavy and very light Majorana neutrinos, the relation between
fermion and gauge boson mixing angles, etc., are just consequences of
mathematical consistency requirements of the proposed SU(3) conformal
unification scheme of strong and electroweak forces.
The universal ultraviolet cutoff (minimal universal scale or distance) is fixed
by the mass of weak gauge bosons \cite{Palle1,PDG}

\begin{equation}
\Lambda = \frac{2 \pi}{\sqrt{6} g_{w}} M_{W},\ 
g_{w} = e / \sin \Theta_{W},\ 
\alpha_{e} = \frac{e^{2}}{4 \pi}
\Longrightarrow  \Lambda = 321.3 GeV .
\end{equation}

It was shown that heavy Majorana neutrinos could be perfect candidate
particles for cold dark matter \cite{Palle2} because they are cosmologically
stable owing to the absence of Higgs scalars in the theory. They are probably
already indirectly observed in the center of our galaxy by atmospheric 
\v Cerenkov
telescopes \cite{Whipple}.
The universality of the minimal scale is confirmed by the Einstein-Cartan
quantum theory of gravity that can resolve cosmological problems without
inflaton scalar fields \cite{Palle3}. The prediction of the Einstein-Cartan
cosmology for the negative cosmological constant (or the negative contribution
of torsion and zero cosmological constant)
can explain the low
power of the large-angle CMBR data by the integrated Sachs-Wolfe effect if the 
Hubble constant is small (or large).
 If the total angular momentum of the Universe is large
at present then the Hubble constant could be also large explaining 
the anomalous large scale flows of the Universe \cite{PalleF}.
The rotating Universe is a natural consequence of the Einstein-Cartan
cosmology with spinning hot and cold dark matter particles \cite{PalleV}.
This vorticity
can be studied by CMBR (WMAP)\cite{WMAP} or by SDSS data \cite{SDSS}.

Having not only renormalizable theory \cite{Palle4}, but also ultraviolet
finite gauge theory to describe the world of particle physics,
one is faced with a possibility to resolve the complete structure of
all Green functions of the theory studying the respective Dyson-Schwinger
equations.
We start this difficult task in the present paper by the study of
Dyson-Schwinger equations for Dirac fermion propagators in the
ladder approximation where
fermions are coupled to one massive gauge boson.
We give the respective equations and necessary algorithms to solve 
these equations
in the next section, while the results and thorough analyses and
discussion are
given in the last section.

\section{Equations and algorithms}
We assume that Dirac fermions couple chirally symmetric
to one massive gauge boson by the standard form (M$\equiv$gauge boson mass),
thus we study the Abelian chirally symmetric version of the BY theory
of the Ref.\cite{Palle1}:

\begin{eqnarray}
{\cal L} &=& \overline{\Psi}_{D}(\imath \not\partial
- g \not A )\Psi_{D}  - \frac{1}{4} F_{\mu\nu}
F^{\mu\nu} + {\cal L}_{g.f.} \nonumber \\
& &+(\partial_{\mu}\Phi^{*}+\imath g A_{\mu}\Phi^{*})
(\partial^{\mu}\Phi-\imath g A^{\mu}\Phi)-
(Y_{M}\overline{\Psi}_{D}\Phi^{*}\Psi^{c}_{D}+h.c.), \\
{\cal L}_{g.f.}&=&standard\ gauge\ fixing\ terms,\ 
\Psi^{c}_{D}\equiv charge\ conjugated\ \Psi_{D}, \nonumber \\
F_{\mu\nu}&=&\partial_{\mu}A_{\nu}-\partial_{\nu}A_{\mu},\ 
\Phi = v+\imath \chi,\ M=\sqrt{2} g v. \nonumber
\end{eqnarray}

The renormalizability and gauge invariance are ensured by
the coupling of the Nambu-Goldstone boson to the gauge boson and the Majorana
fermion as in \cite{Palle1}. We assume also that Nambu-Goldstone bosons carry
lepton number as in the BY theory of Ref.\cite{Palle1}, thus only Majorana
bare mass term is allowed. The symmetry breaking parameter $v$ is in this
model free parameter and it is not fixed by the Wick's theorem as in
the non-Abelian version BY of \cite{Palle1}.
We study only equations of a Dirac fermion
in this paper.

The Dirac fermion propagator is defined as
$ S'_{F}(p)=[\alpha(p)\not p-\beta(p)]^{-1}$.
It is advantageous to write Dyson-Schwinger equations for
fermion propagators in the ladder approximation \cite{Fukuda},
in the Landau gauge. Then, the equations have the following
form in the spacelike domain:

\begin{eqnarray}
\beta (x) &=& C \int^{\Lambda^{2}}_{0} dy K(x,y)
\frac{y \beta (y)}{y\alpha^{2}(y)+\beta^{2}(y)},\nonumber \\ 
\alpha (x) &=& 1 - C \int^{\Lambda^{2}}_{0} dy L(x,y)
\frac{y \alpha (y)}{y\alpha^{2}(y)+\beta^{2}(y)}, \\
x &\equiv & p^{2},\ C \equiv \alpha_{g}/\pi \equiv \frac{g^{2}}{4\pi^{2}}, 
\ 
K(x,y) = \frac{3}{2} (1 + \frac{M^{2}}{3s})\frac{1}{x+y+M^{2}+s}, 
\nonumber \\
L(x,y) &=& \frac{y M^{2}}{s (x+y+M^{2}+s)^{2}},
\ s = [(x+y+M^{2})^{2}-4 x y ]^{1/2}. \nonumber
\end{eqnarray}

One has to add one more term in the timelike domain because
of the existence of the branch point of kernels and a correct analytical 
continuation \cite{Fukuda,Palle4}:

\begin{eqnarray}
\beta (x) &=& C \int^{\Lambda^{2}}_{0} dy K(x,y)
\frac{y \beta (y)}{y\alpha^{2}(y)+\beta^{2}(y)} \nonumber \\
&+& \Theta(\sqrt{-x}-M) C \int^{0}_{-(\sqrt{-x}-M)^{2}}
dy \Delta K (x,y)\frac{y \beta (y)}{y\alpha^{2}(y)+\beta^{2}(y)},                      \nonumber \\
\alpha (x) &=& 1 - C \int^{\Lambda^{2}}_{0} dy L(x,y)
\frac{y \alpha (y)}{y\alpha^{2}(y)+\beta^{2}(y)} \nonumber \\
&-& \Theta(\sqrt{-x}-M) C \int^{0}_{-(\sqrt{-x}-M)^{2}}
dy \Delta L (x,y)\frac{y \alpha (y)}{y\alpha^{2}(y)+\beta^{2}(y)}, 
\end{eqnarray}
\begin{eqnarray*}
s = \imath t,\ t^{2} = 4xy-(x+y+M^{2})^{2},\ 
\Delta K \equiv K(x,y,s)-K(x,y,s^{*}),\\ 
\Delta L \equiv L(x,y,s)-L(x,y,s^{*})  \\
\Rightarrow 
\Delta K(x,y) = - \frac{3s}{4xy}+\frac{M^{2}(x+y+M^{2})}{4xys},\\ 
\Delta L(x,y) = \frac{M^{2}}{4yx^{2}s}[(x+y+M^{2})^{2}-2xy]. 
\end{eqnarray*}

Now we shall describe in detail how we solve equations, while anayses
of solutions are left for the final section.

We solve equations in four steps.

{\bf Step 1}: 

 The equations in the spacelike domain have the
form of the coupled nonlinear Fredholm integral equations of the
second kind \cite{Baker} and we need the initial guess functions
to proceed further.
A good choice is a solution of the nonlinear equations for
the vanishing gauge boson mass. In this case, the nonlinear 
integral equations are reduced to nonlinear differential 
equations \cite{Fukuda}:

\begin{eqnarray}
(4x\frac{d^{2}}{dx^{2}} + 8\frac{d}{dx})
 B(x) = -3C\frac{B(x)}{x+B^{2}(x)},\ \beta=B,\ \alpha=1.
\end{eqnarray}

This equation is solved by the Adams-Bashforth method. One can easily find
initial conditions from the equation and using the rule of de l'H\^{o}pital
at $x=p^{2}=0$:

\begin{eqnarray*}
\frac{dB}{dx} (0) = -\frac{3}{8}\frac{C}{B(0)},\ 
\frac{d^{2}B}{dx^{2}}(0) = -\frac{3C^{2}}{32 B^{3}(0)}.
\end{eqnarray*}

The asymptotic of the solution at large spacelike momenta gives us 
a condition to find the solution with an arbitrary number of nodes

\begin{eqnarray*}
\frac{dB}{dx}(x=\Lambda_{i}^{2}) + \frac{B(x=\Lambda_{i}^{2})}
{\Lambda_{i}^{2}}=0.
\end{eqnarray*}

One can simply generate solutions from fixed initial conditions at
the zero momentum and then search for momenta which fulfill
above condition. As an example, for $C=0.7$ and $B(0)=1 GeV$, one
obtains $j$-node solutions with the following $\Lambda_{j}$
cutoffs: $\Lambda_{0}=4.292 GeV$, $\Lambda_{1}=85.82 GeV$ 
and $\Lambda_{2}=1715.87 GeV$.

Thus we can generate initial guess functions by $\beta=B$ and $\alpha=1$.
Assuming the existence of the fundamental cutoff $\Lambda$, one has
to rescale all dimensional quantities of a certain dimension $k$ by
multiplication with $(\Lambda/\Lambda_{i})^{k}$ at the end of the calculation
with the cutoff $\Lambda_{i}$.

{\bf Step 2}: 

Using the initial guess functions, we have to choose the method how to
solve nonlinear integral equations. Between Nystr\"{o}m, Galerkin or
 the collocation method \cite{Baker}, or Newton-like iterations \cite{Salanova},
we decide to implement the collocation method.

It is more comfortable to work with logarithmic variables in the
spacelike domain, thus we change the variables to $w=\ln (1+x/B(0)^{2})$.
The $\beta$ and $\alpha$ functions are approximated by
\v Cebi\v sev polynomials:

\begin{eqnarray*}
f(x) \approx -\frac{1}{2}c_{1} + \sum^{N}_{k=1}c_{k}T_{k-1}(x),
\ -1 \leq x \leq +1.
\end{eqnarray*}

Inserting $\tilde{\beta}$ and $\tilde{\alpha}$ into the Fredholm
equations, Eq.(3), we obtain a nonlinear algebraic system of equations for the
set of coefficients $\{b_{i},a_{j}\}$ in Eq. (6).
This system of equations is solved by the modified Powell hybrid method
\cite{Powell}.

We divide the integration region for a one-node solution into two segments
$ 0 \leq x \leq \Lambda^{2}_{0}$ and  $ \Lambda^{2}_{0} \leq x 
\leq \Lambda^{2}_{1} $, and write the approximate solutions
$\tilde{\beta}$ and $\tilde{\alpha}$ with two separate \v Cebi\v sev
expansions, one for every segment.
 The accuracy of the approximation is
improved, but the number of variables is twice compared with the zero-node
case.Usually, we take 20 \v Cebi\v sev polynomials in one approximation,
thus $2\times 2\times 20=80$ coefficients as variables for 
one-node solution, 40 for zero-node, 160 for two-node solution, etc.
We define the discretized sequence of the corresponding 
logarithmic variables $w_{i}$ in each segment that is 
distributed homogeneously, except for a denser distribution 
of points at the joint of two segments.
The corresponding number of nonlinear algebraic equations
for coefficients $b_{i},\ a_{j}$ can now be formed:

\begin{eqnarray*}
F_{1,i} &\equiv & \tilde{\beta}(w_{i})-CB(0)^{2}\int^{w_{\Lambda}}_{0}
dw' e^{w'}K(x(w_{i}),y(w'))
\frac{y(w')\tilde{\beta}(w')}{y(w')\tilde{\alpha}^{2}(w')+
\tilde{\beta}^{2}(w')} = 0, \\
F_{2,i} &\equiv & \tilde{\alpha}(w_{i})-1+CB(0)^{2}\int^{w_{\Lambda}}_{0}
dw' e^{w'}L(x(w_{i}),y(w'))
\frac{y(w')\tilde{\alpha}(w')}{y(w')\tilde{\alpha}^{2}(w')+
\tilde{\beta}^{2}(w')} = 0,
\end{eqnarray*}
\begin{eqnarray}
i = 1,...,n,
\end{eqnarray}
\begin{eqnarray*}
\tilde{\beta}(w)&=&\sum^{n}_{k=1}b_{k}T_{k-1}(\overline{w}),\ 
\tilde{\alpha}(w)=\sum^{n}_{k=1}a_{k}T_{k-1}(\overline{w}),\ 
\overline{w}=\frac{2w}{w_{\Lambda}}-1, \\
w_{\Lambda}&=&\ln (1+\Lambda^{2}/B(0)^{2}),\ x(w)=B(0)^{2}(e^{w}-1),
\ y(w)=B(0)^{2}(e^{w}-1). 
\end{eqnarray*}

To this system, we apply the modified Powell hybrid method and verify 
the result of the computation.

{\bf Step 3}:

If we obtain a solution even under small tolerance, one has 
to verify it with a very large number of arguments $w$
(more than 1000) for both $\tilde{\beta}$ and $\tilde{\alpha}$
and check the errors. One can expect larger errors for higher-node
solutions.

The final check whether our solution is a real solution or only some
local minimum must be performed by solving the system once more,
but now with a more precise approximation scheme, namely, the piecewise cubic
spline method.

The variables are now values of $\tilde{\beta}$ and $\tilde{\alpha}$
functions evaluated for 200 arguments $w_{i}$ and we form equations
as was done previously for \v Cebi\v sev coefficients.
 Thus, we solve an
algebraic system with 400 variables using the modified Powell
hybrid method. Note that for every solution $(\beta, \alpha)$
there is also a solution $(-\beta, \alpha)$.
If the procedure does not diverge or does not end with a trivial solution,
we have a very accurate solution in the spacelike domain that
should be checked explicitly once more.

{\bf Step 4}:

It is obvious that the relations in Eq.(4) are not equations
for the timelike domain, but just formulas for evaluations
of $\beta$ and $\alpha$ in the timelike domain, knowing their
solutions in the spacelike domain.

Namely, let us define the following intervals:
$I_{0}=\{ x | 0 \leq x \leq \Lambda^{2}\}$ and
$I_{j}=I_{0} \bigcup \{ x | 0 \geq x \geq -(j\times M)^{2} \}$.
From the boundaries of integrals in Eq.(4) one can conclude
that the timelike part of $I_{1}$ is defined by integration 
over $I_{0}$, similarly, the timelike part of $I_{2}$ is defined
by integration over $I_{1}$, etc.
The procedure can be continued to arbitrary $I_{k}$.

Note that the integrals over timelike domains are very well 
defined by the endpoint singularity integration. The endpoint
singularity at $y=-(\sqrt{x}-M)^{2}$ of kernels $\Delta K$
and $\Delta L$ is integrable.

If the propagator reaches the mass singularity for a certain
$x = -m_{f}^{2} \in I_{k}$, then at $x = -(m_{f}+M)^{2} \in
 I_{k+1}$ the functions $\beta$ and $\alpha$ diverge because
the endpoint singularity becomes nonintegrable. However,
the mass function $\mu (x) \equiv \beta (x) / \alpha (x)$
is even at this point well defined as the limes of the
quotient $\infty / \infty $. 
Anyhow, the numerics in the interval
$I_{k+1}$ in the vicinity of $x = -(m_{f}+M)^{2}$ is becoming
difficult, so we stop our calculation at $I_{k}$.

We perform integrations using the Gauss-Kronrod and Clenshaw-Curtis methods
which we have checked using the slow and accurate Monte Carlo method.
The functions $\beta$ and $\alpha$ are approximated in the timelike
part of $I_{k}$ by piecewise cubic spline with 800 values for each
function.

This completes our procedure of solving Dyson-Schwinger equations
for fermion propagators. The next section is devoted to results and
physical implications.

\section{Results and discussion}

The results presented in this paper are only those that have passed
our procedure in four steps. The fact that the procedure fails 
in certain cases does not mean that the equations do not necessarily 
have solutions. The proof of the existence of any kind of solutions 
is not attempted in this paper.

Our experience with these equations tells us that our procedure 
fails to find solutions for higher gauge boson mass, one-node
solutions for coupling close to the critical $C=\frac{1}{3}$,
and two- or higher-node solutions for any coupling.
However, we find enough solutions to make relevant physical
conclusions and suggestions for the improvement of algorithms.
The attempt to find a
 solution by perturbing parameters, such as
coupling or boson mass, is usually not successful.

The reader can visualize solutions with zero and one node in
Figs.1-4, where the mass function is defined as 
$\mu \equiv \beta / \alpha$ and $m_{f}\equiv$ fermion mass.

\begin{figure}
\epsfig{figure=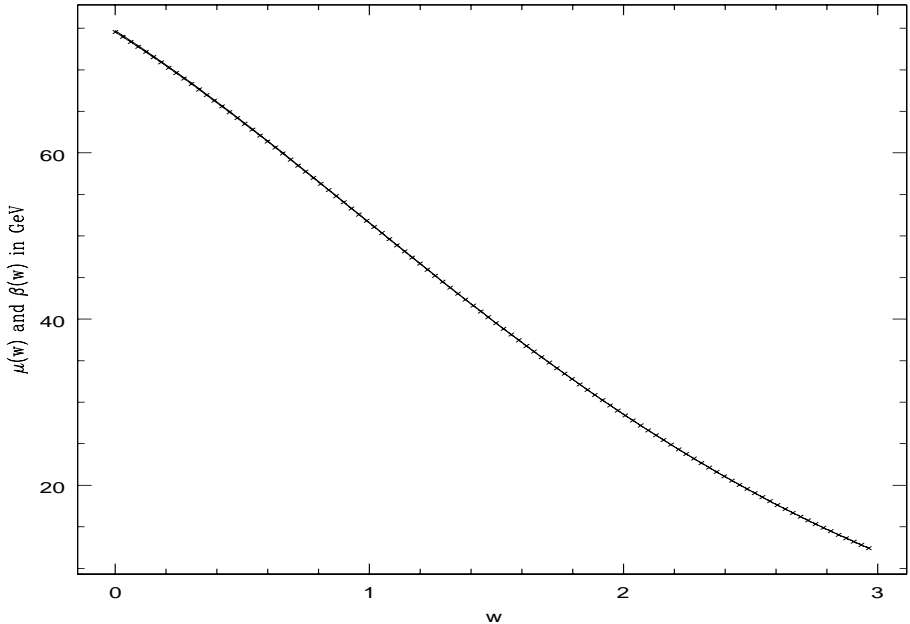, height=90 mm, width=130 mm}
\caption{This figure depicts the mass-function $\mu$ (solid line)
and the $\beta$-function (crosses) in the spacelike domain for
a zero-node solution and parameters $C=0.7,\ M=3.5 GeV,\ 
\Lambda=321.3 GeV$. "w" is a logarithmic variable defined by
$B(0)=1 GeV$.}
\end{figure}

\begin{figure}
\epsfig{figure=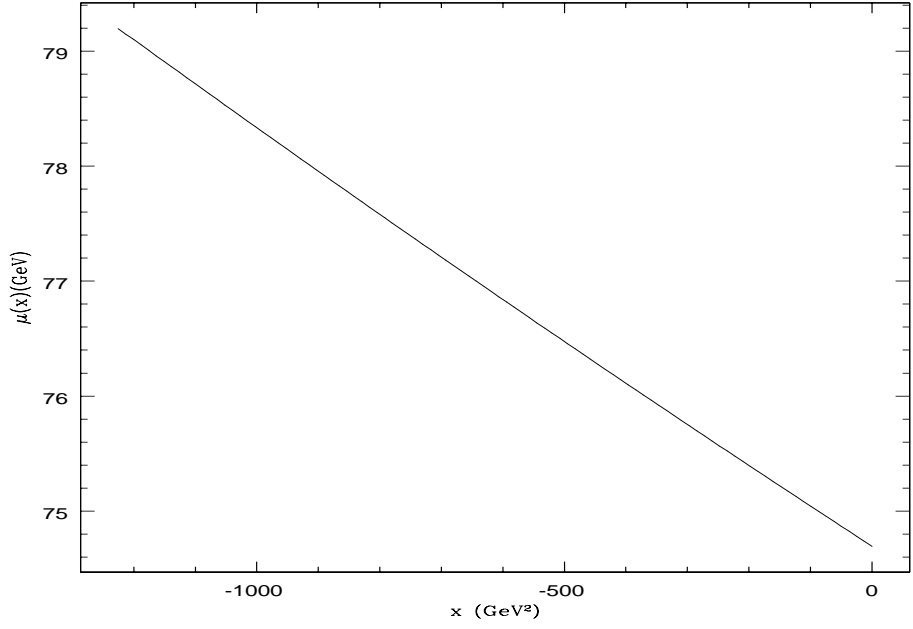, height=90 mm, width=130 mm}
\caption{This figure depicts the mass-function $\mu$ 
 in the timelike domain for
a zero-node solution and parameters $C=0.7,\ M=3.5 GeV,\ 
\Lambda=321.3 GeV$.}
\end{figure}

\begin{figure}
\epsfig{figure=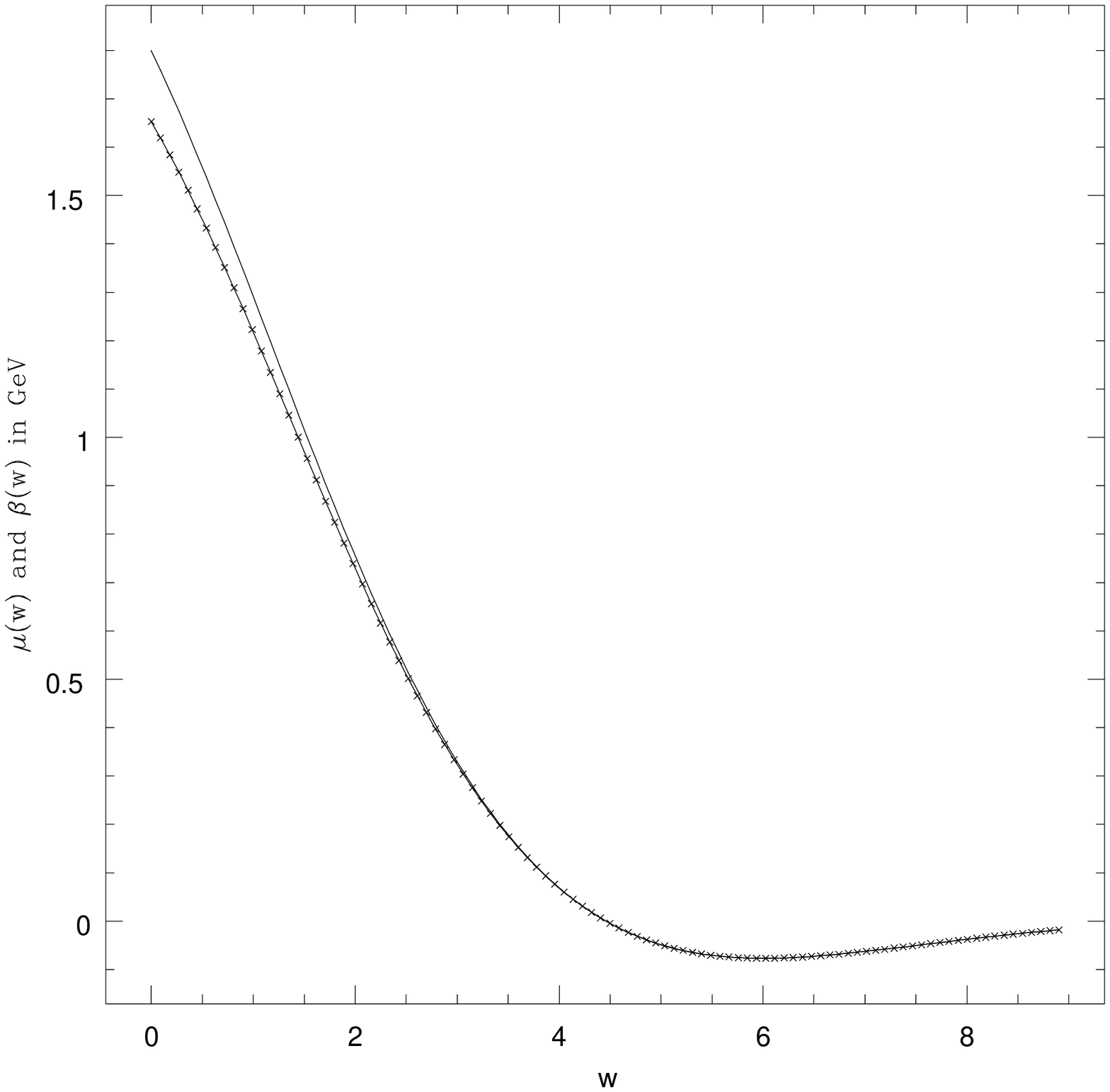, height=90 mm, width=130 mm}
\caption{The mass-function $\mu$ (solid line)
and the $\beta$-function (crosses) in the spacelike domain for
a one-node solution and parameters $C=0.7,\ M=3.5 GeV,\ 
\Lambda=321.3 GeV$. "w" is a logarithmic variable defined by
$B(0)=1 GeV$.}
\end{figure}

\begin{figure}
\epsfig{figure=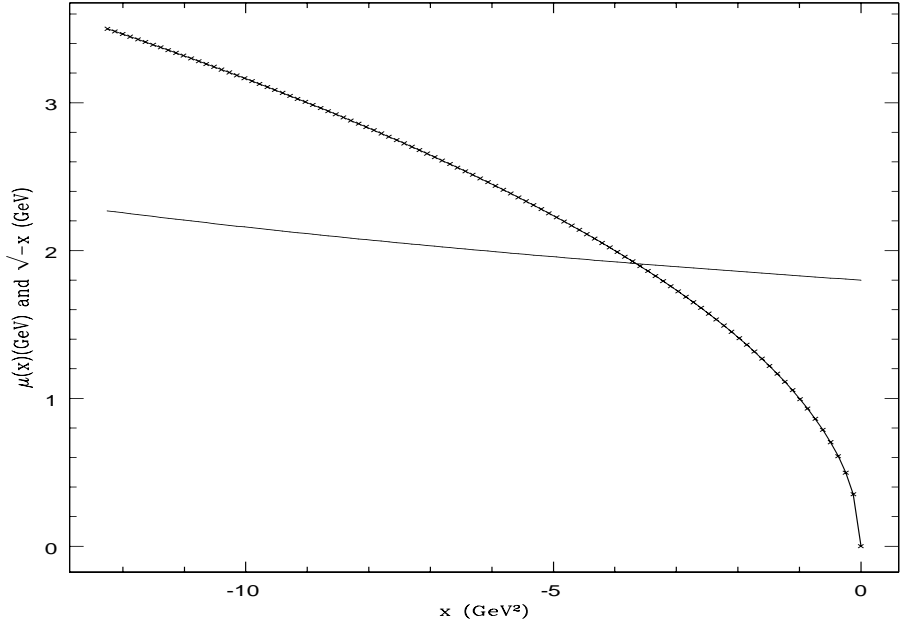, height=90 mm, width=130 mm}
\caption{The mass-function $\mu(x)$ (solid line) and
$\sqrt -x$ (crosses) 
 plotted in the timelike domain for
a one-node solution and parameters $C=0.7,\ M=3.5 GeV,\ 
\Lambda=321.3 GeV$.}
\end{figure}

\begin{figure}
\epsfig{figure=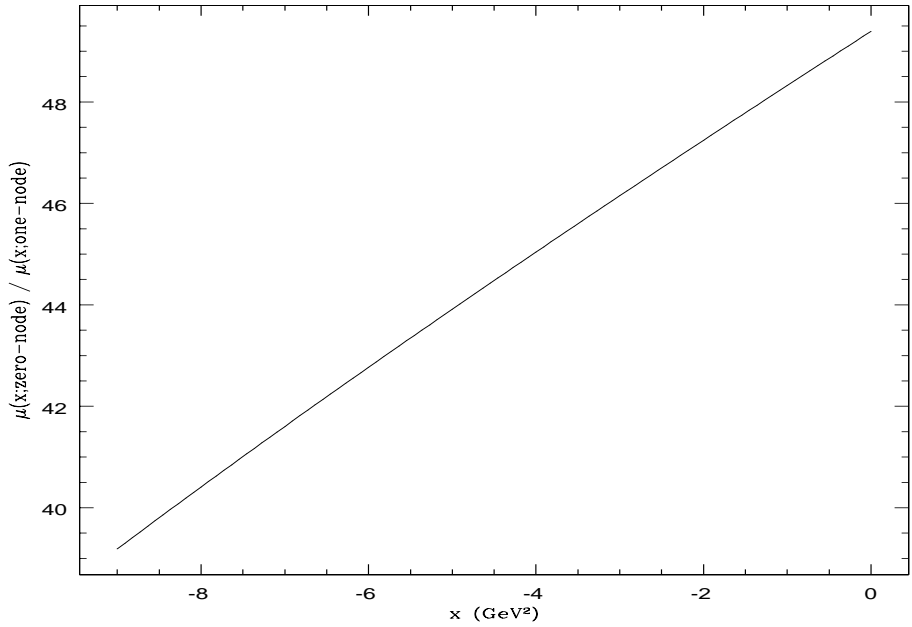, height=90 mm, width=130 mm}
\caption{The quotient of the mass-functions $\mu$ of
 zero-node and one-node solutions
 in the timelike domain for
 parameters $C=0.75,\ M=3.0 GeV,\ 
\Lambda=321.3 GeV$.}
\end{figure}

\begin{figure}
\epsfig{figure=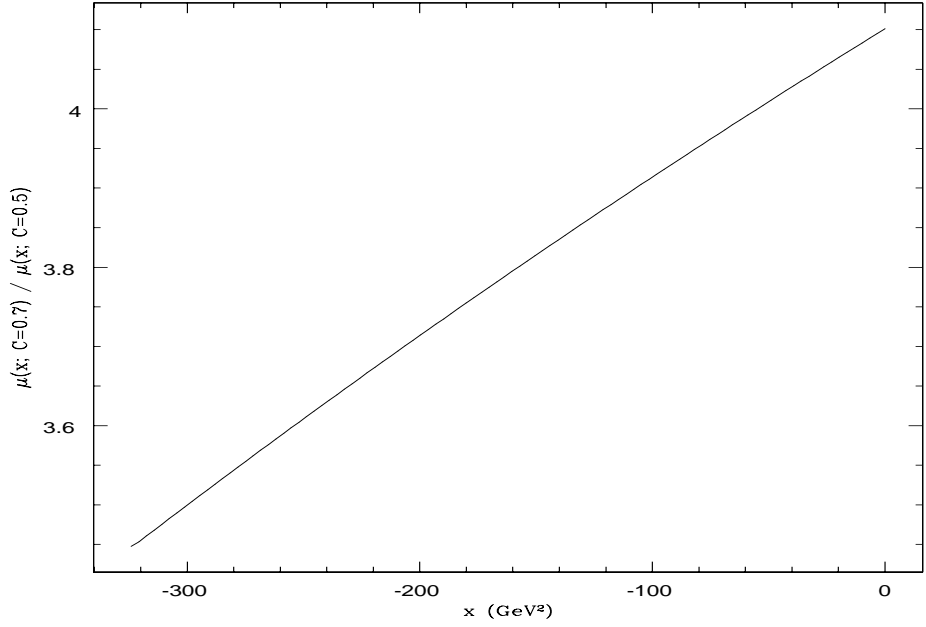, height=90 mm, width=130 mm}
\caption{The quotient of the mass-functions $\mu$ 
of zero-node solutions for $C=0.7$ and $C=0.5$
 in the timelike domain and parameters $M=4.5 GeV,\ 
\Lambda=321.3 GeV$.}
\end{figure}

That zero-node solutions are
heavier than one-node solutions can be seen from Figs. 1-5.
Equations (3) have solutions only in the strong coupling
regime $C\equiv \alpha_{g}/\pi > \frac{1}{3}$.
Verifying the asymptotics of the zero-node solutions in the
timelike domain,
one can conclude on the absence of the mass singularity for 
couplings or gauge-boson masses when we have nontrivial 
solutions. This is a consequence of the strong coupling between
fermions and the gauge boson. Usually, the one-node solutions have
mass singularity (see Tables 1-3), unless one reaches a very strong
coupling regime.

\begin{table}
\begin{tabular}{|c||c|c|c|c|c|c|} \hline
$M(GeV)$ & 2.0 & 2.5 & 3.0 & 3.5 & 4.0 & 4.5 \\ \hline
$m_{f}(GeV)$ & - & 4.932 & 2.528 & 1.912 & 1.310 & 0.642 \\
\hline
\end{tabular}
\caption{
The dependence of $m_{f}$ on the gauge boson mass M
for a one-node solution with $\Lambda=321.3 GeV$ and $C=0.7$;
sign "-" denotes the absence of the mass singularity}
\end{table}

\begin{table}
\begin{tabular}{|c||c|c|c|c|c|} \hline
$C$ & 0.65 & 0.675 & 0.70 & 0.725 & 0.75 \\ \hline
$m_{f}(GeV)$ & 0.351 & 1.401 & 2.528 & 5.833 & - \\ \hline
\end{tabular}
\caption{
The dependence of $m_{f}$ on the coupling constant $C$
for a one-node solution with $\Lambda=321.3 GeV$ and $M=3.0 GeV$
}
\end{table}

\begin{table}
\begin{tabular}{|c||c|c|c|c|c|} \hline
$\Lambda(GeV)$ & 214.2 & 241.0 & 275.4 & 321.3 & 482.0 \\ \hline
$m_{f}(GeV)$ & 0.428 & 0.983 & 1.639 & 2.528 & - \\ \hline
\end{tabular}
\caption{
The behavior of $m_{f}$ on the fundamental cutoff $\Lambda$
for a one-node solution with $M=3 GeV$ and $C=0.7$
}
\end{table}

\begin{table}
\begin{tabular}{|c||c|c|c|c|c|c|c|} \hline
$M(GeV)$ & 6.0 & 8.0 & 10.0 & 12.0 & 14.0 & 16.0 & 18.0 \\ \hline
$m_{f}(GeV)$ & - & 22.43 & 15.27 & 23.90 & 9.84 & 7.24 & 4.44 \\ \hline
\end{tabular}
\caption{
The behavior of $m_{f}$ on the gauge boson mass $M$
for a zero-node solution with $\Lambda=321.3 GeV$ and $C=0.5$ 
in the "model" case $\alpha\equiv 1$
}
\end{table}

\begin{table}
\begin{tabular}{|c||c|c|c|c|c|c|} \hline
C  &  0.65 & 0.65 & 0.70 & 0.70 &
 0.75 & 0.75  \\ \hline
n & 0 & 1 & 0 & 1 & 0 & 1 \\ \hline
$b_{UV}$ & -1.003 & -1.0015 & -1.0021 & -1.00937 & -1.0014 & -1.0076 \\
\hline
\end{tabular}
\caption{
The UV asymptotic index $b_{UV}$ for various couplings C, M=3 GeV
and $\Lambda=321.3 GeV$; n=number of nodes. 
}
\end{table}

The "model" case when $\alpha \equiv 1$ is studied for zero-node
solutions with 
results in Table 4, where one can see the appearence of the 
mass singularity and its behavior on the gauge boson mass.
The same "model" is studied in \cite{Cheng} but it is deficient because
of various reasons: (1) it is not gauge invariant within a ladder
approximation, (2) nontrivial solutions emerge only in the strong
coupling regime, so $\alpha$ function affects crucially the mass function and
cannot be set to 1, (3) according to the discussion in Step 4 of
the description of the solving-procedure it is evident that the
mass function contains spurious unphysical singularity in the segment
$I_{k+1}$ in the timelike domain if we put $\alpha \equiv 1$.

The behavior of the mass of the one-node solutions on the
gauge boson mass can be read in Table 1. Fermion masses are
very sensitive to the boson mass and they are larger for smaller
boson masses. This is the expected feature, because the fermion
mass singularity disappears for the vanishing boson mass 
\cite{Fukuda}.

Larger coupling implies larger fermion self-energy, thus also
larger fermion masses, as one can read from Table 2.
By rescaling of cutoffs and masses in Table 1, in Table 3 we show 
the dependence of fermion masses on the fundamental
cutoff for the fixed boson mass. The guess that larger cutoff
means larger self-energy and consequently larger masses is
completely confirmed.

Similar dependencies are valid for zero-node fermion states.
We depict some comparisons of mass functions in Fig.6.

Let us define infrared (IR) and ultraviolet (UV) asymptotics
of the mass function:

\begin{eqnarray*}
\mu (p^{2}) \sim  (p^{2})^{b}:\ 
b_{IR}=\lim_{x \rightarrow 0} \frac{x d\mu(x)/dx}{\mu(x)},\ 
b_{UV}=\lim_{x \rightarrow \Lambda^{2}} \frac{x d\mu(x)/dx}{\mu(x)}.
\end{eqnarray*}

A direct inspection of our solutions leads to the following 
conclusions: (1) $b_{IR}=0$ for $M=0$ and $M\neq 0$,
(2) $b_{UV}=-1$ for $M=0$ and there is a small deviation from -1 for
$M\neq 0$ (see Table 5).

We can now make final conclusions about solutions of Dyson-Schwinger
equations. We have shown that even the most simple bootstrap system
studied in this paper has at least two solutions for the fixed coupling,
the gauge boson mass, and the cutoff. The heaviest solution is the 
zero-node solution (belonging to the third fermion family in 
the standard particle physics classification),
while the lighter one is the one-node solution (belonging to the
second fermion family). The mass gap between these two solutions
is of the order of magnitude observed in nature (Figs. 1-5).
 We need more sophisticated approximation
schemes for functions to search for higher-node solutions,
thus to find the first-family member fermion.
The observed behavior of the mass functions on the parameters
fulfills our expectations.

One should consider our work as the starting attempt to solve 
the family replication problem which is not soluble with the Higgs mechanism.
Two basic scenarios are possible: (1) there are only three physical
solutions in the spacelike domain, (2) there are more than three
solutions in the spacelike domain but only three physically
acceptable also within the timelike domain. Future studies will answer
which scenario will prevail.

Before turning to a more complicated electroweak system,
we are expecting to see the verification of the principle
of noncontractibility by the LHC. The nonresonant enhancement of
the QCD amplitudes at the weak scale has been reported
by the TeVatron \cite{Palle4}, especially measuring and estimating
the quotient of cross sections at two center of mass energies 
\cite{TeVatron,Palle5}. High-luminosity measurements at the LHC could
determine the ultraviolet cutoff to high accuracy.


\begin{thebibliography}{100}

\bibitem{Palle1} D. Palle, Nuovo Cimento {\bf A 109}, 1535 (1996)
\bibitem{PDG} W. M. Yao et al., 
              J. Phys. {\bf G 33}, 1 (2006)
\bibitem{Palle2} D. Palle, Nuovo Cimento {\bf B 115}, 445 (2000);
                 D. Palle, ibidem {\bf B 118}, 747 (2003)
\bibitem{Whipple} 
  F. Aharonian et al.,  
 Astron. and Astrophys. {\bf 425}, L13 (2004);
 J. Albert et al.,
  Astrophys. J {\bf 638}, L101 (2006)
\bibitem{Palle3} D. Palle, Nuovo Cimento {\bf B 111}, 671 (1996);
                 D. Palle, ibidem {\bf B 114}, 853 (1999)
\bibitem{PalleF} D. Palle, preprint {\bf arXiv:0902.1852} (2009) 
\bibitem{PalleV} D. Palle, preprint {\bf arXiv:0802.2060} (2008)
\bibitem{WMAP} K. Land K, J. Magueijo, Phys. Rev. Lett. {\bf 95},
                 071301 (2005)
\bibitem{SDSS} M. J. Longo, preprint {\bf arXiv:0812.3437} (2008)
\bibitem{Palle4} D. Palle, Hadronic J. {\bf 24}, 87 (2001);
                 D. Palle, ibidem {\bf 24}, 469 (2001)
\bibitem{Fukuda} T. Maskawa, H. Nakajima, Prog. Theor. Phys. {\bf 52},
  1326 (1974); T. Maskawa, H. Nakajima, ibidem {\bf 54}, 860 (1975);
  R. Fukuda, T. Kugo, Nucl. Phys. {\bf B 117}, 250 (1976)
\bibitem{Baker} C. T. H. Baker, {\sl The Numerical Treatment of
  Integral Equations}, (Clarendon Press, Oxford 1977);
  K. E. Atkinson, {\sl The Numerical Solution of Integral Equations
  of the Second Kind}, (Cambridge University Press, Cambridge 1997)
\bibitem{Salanova} M. A. Hernandez, M. A. Salanova,
  J. Integral Equ. and Applic. {\bf 17}, 1 (2005)
\bibitem{Powell} M. J. D. Powell, in {\sl Numerical methods for
   Nonlinear Algebraic Equations, ed. by P. Rabinowitz, p.87},
   (Gordon and Breach, London, 1970)
\bibitem{Cheng} G. Cheng, T. K. Kuo, J. Math. Phys. {\bf 35}, 6270 (1994)
\bibitem{TeVatron} B. Abbott et al. 
      Phys. Rev. Lett. {\bf 86} 2523 (2001);
  T. Affolder, et al.,
    ibidem {\bf 88}, 042001 (2002)
\bibitem{Palle5} D. Palle, preprint {\bf arXiv:0910.3852} (2009)

\end{thebibliography}
\end{document}